\documentclass[fleqn,12pt]{wlscirep}
\usepackage[font=footnotesize,labelfont=bf]{caption}
\usepackage[right]{lineno}
\usepackage{hyperref}
\hypersetup{colorlinks=true, linkcolor=cyan, filecolor=cyan, urlcolor=cyan}

\usepackage{lettrine}
\graphicspath{{Figures/}{../Figures/}}
\usepackage[textsize=tiny]{todonotes}

\usepackage{sectsty}
\definecolor{heading_color}{HTML}{a00000}
\sectionfont{\color{heading_color}}  

\title{Antifragility as a complex system's response to perturbations, volatility, and time}

\author[1]{Cristian Axenie}
\author[2]{Oliver L\'opez-Corona}
\author[3]{Michail A. Makridis}
\author[4]{Meisam Akbarzadeh}
\author[5]{Matteo Saveriano}
\author[6]{Alexandru Stancu}
\author[7,*]{Jeffrey West}

\affil[1]{Department of Computer Science and Center for Artificial Intelligence,
Nuremberg Institute of Technology Georg Simon Ohm, Nuremberg, Germany}
\affil[2]{Investigadores por M\'exico (IxM) at Instituto de Investigaciones en Matem\'aticas Aplicadas y Sistemas (IIMAS), Universidad Nacional Autónoma de México (UNAM), Ciudad Universitaria, CDMX, M\'exico}
\affil[3]{IVT, Civil Environmental and Geomatic Engineering, ETH Zurich, Switzerland}
\affil[4]{Department of Transportation Engineering, Isfahan University of Technology, Isfahan, Iran}
\affil[5]{Department of Industrial Engineering, University of Trento, Trento, Italy}
\affil[6]{Department of Electrical and Electronic Engineering, The University of Manchester, Manchester, UK}
\affil[7]{Department of Integrated Mathematical Oncology,
H. Lee Moffitt Cancer Center \& Research Institute, Tampa, FL, USA \vspace{0.5cm}}
\affil[*]{jeffrey.west@moffitt.org\vspace{0.5cm}}

\begin{abstract}
Antifragility characterizes the benefit of a dynamical system derived from the variability in environmental perturbations. Antifragility carries a precise definition that quantifies a system's output response to input variability. Systems may respond poorly to perturbations (fragile) or benefit from perturbations (antifragile). In this manuscript, we review a range of applications of antifragility theory in technical systems (e.g., traffic control, robotics) and natural systems (e.g., cancer therapy, antibiotics). While there is a broad overlap in methods used to quantify and apply antifragility across disciplines, there is a need for precisely defining the scales at which antifragility operates. Thus, we provide a brief general introduction to the properties of antifragility in applied systems and review relevant literature for both natural and technical systems' antifragility. We frame this review within three scales common to technical systems: intrinsic (input-output nonlinearity), inherited (extrinsic environmental signals), and interventional (feedback control), with associated counterparts in biological systems: ecological (homogeneous systems), evolutionary (heterogeneous systems), and interventional (control). We use the common noun in designing systems that exhibit antifragile behavior across scales and guide the reader along the spectrum of fragility--adaptiveness--resilience--robustness--antifragility, the principles behind it, and its practical implications.
\end{abstract}

\begin{document}
\maketitle
\thispagestyle{empty}

\section{Introduction}
\begin{figure}
    \centering
    \includegraphics[width=1\linewidth]{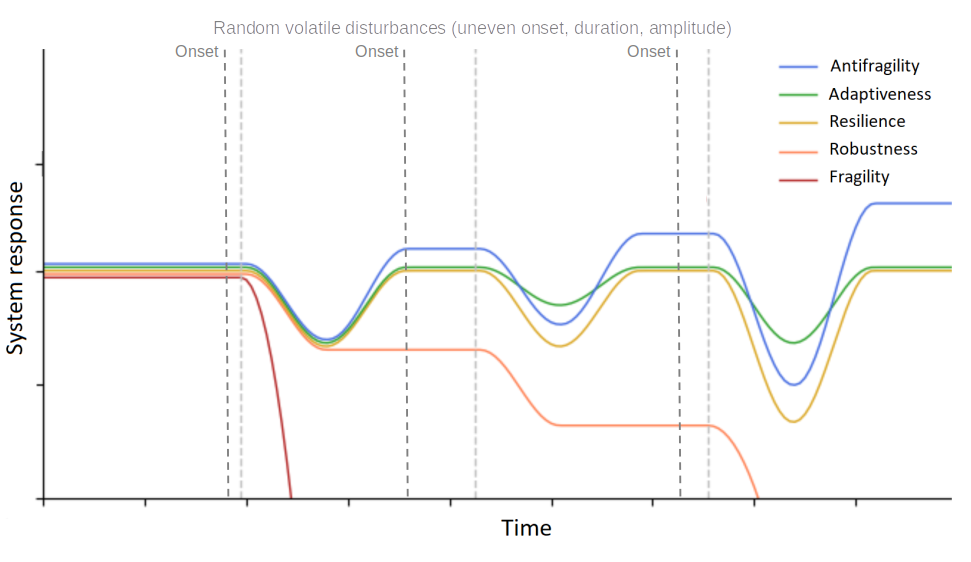}
    \caption{Figure adapted from Axenie et Saveriano\cite{axenie2023antifragilerobotics} - Perspective on a dynamical system's behaviors spectrum: Fragile--Adaptive--Resilient--Robust--Antifragile responses. Example disturbances have random onset, duration (i.e., volatility), and amplitude. Each type of system response captures the predominant traits of the spectrum members.}
    \label{fig:spectrum}
\end{figure}

\lettrine[lines=3]{A}{ntifragile} is a term coined to describe the opposite of fragile, as defined in a recent book that generated significant interest in both the public and scientific domain\cite{taleb2012antifragile}. Although the term has a wide range of applications, it contains a precise and mathematical definition. Systems or organisms can be defined as antifragile if they derive benefit from systemic variability, volatility, randomness, or disorder\cite{taleb2013mathematical}. To get an intuition, we provided samples of a system's reference behaviors in Fig.~\ref{fig:spectrum}, where three small and large disruptions (i.e., in amplitude) occur at random times and inject, through their unexpected onset and duration, volatility, randomness, or disorder in a system's dynamics.

In mathematical terms, antifragility is a nonlinear convex response to a well-defined payoff function that a system exhibits in the face of volatility. This response enhances the system's ability to not only withstand perturbations (robust) but even benefit from them (antifragile). Although the antifragility framework emerged in the context of financial risk analysis, due to its universal mathematical formalism and principles, it has recently drawn attention and has been applied within different concepts across domains: in biology\cite{pineda2018Antifragility}, in socio-economics\cite{de2020antifragility}, in urban planning\cite{blevcic2020antifragile}, and risk analysis\cite{johnson2013antifragility}. Applied antifragility theory extends Nassim Taleb's antifragility principle to the rank of system design methodology across disciplines\footnote{The authors herein have recently formed the Applied Antifragility Group Website: \href{http://antifragility.science/}{https://antifragility.science/}\label{note:antifragilist}}.

\subsection{Defining terms across technical and natural systems}
Herein, we draw from the body of literature on technical systems (e.g., road traffic and robotics control systems) and biological systems (e.g., cancer therapy, antibiotics, and agricultural pest management) to define the scales of antifragility theory and unify definitions from natural and technical systems. The perspective manuscript is organized into three major sections that describe the scales of the applied antifragility spectrum in technical systems: 1) intrinsic, 2) inherited, and 3) induced antifragility. Each scale of antifragility in technical systems has its analogue in natural (biological) systems: ecological, evolutionary, and interventional antifragility. We provide definitions for each proposed antifragility scale and review existing literature to provide examples of successful practical instantiations to a broad audience.

\begin{figure}
    \centering
    \includegraphics[width=1\linewidth]{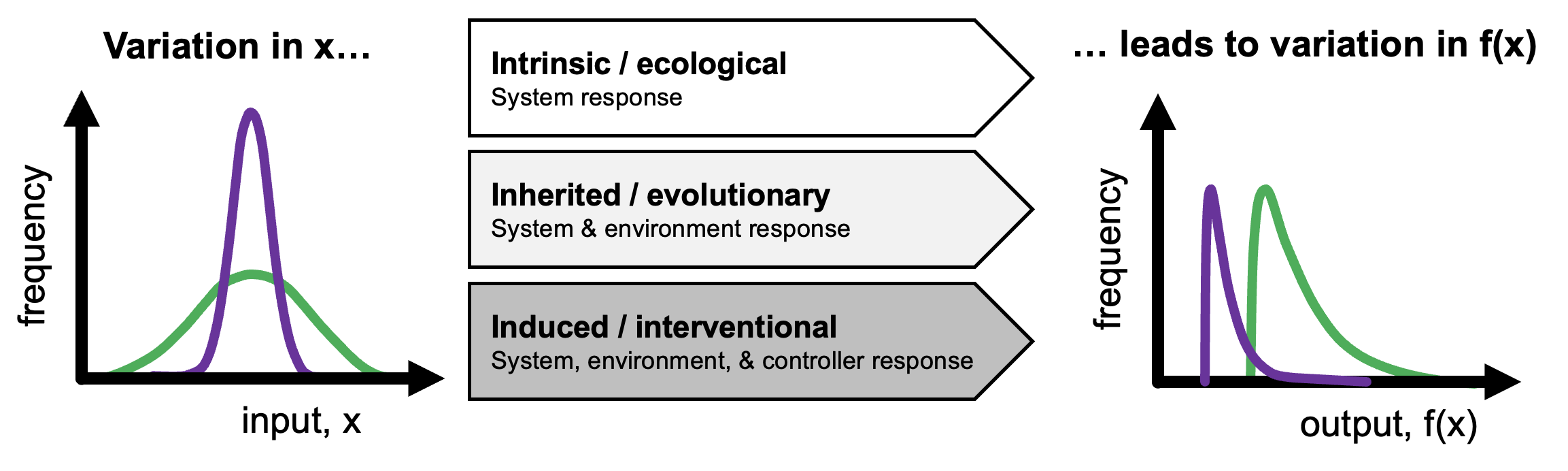}
    \caption{Antifragility-associated terms, defined for technical and natural (biological) systems.}
    \label{table:1}
\end{figure}

As we go up in the spectrum from the bottom (i.e., intrinsic or ecological antifragility), the response properties of the dynamical system and its capacity to gain from stressors and anticipate random and volatile events increase. Intrinsic/ecological antifragility describes the system's internal dynamics without external interactions and intervention. Inherited/evolutionary antifragility describes the system's dynamics under external signals that modulate the system's internal dynamics. Finally, induced/interventional antifragility introduces the closed-loop (behaviour) dimension of antifragility, where the system is connected to control or driving signals. Across these scales, the recently formed Applied Antifragility Group\footref{note:antifragilist} has proposed three \textit{Paths} forward for characterizing, designing, and building systems that behave antifragile in the face of uncertainty, volatility, and randomness:
\begin{itemize}
    \item \textit{Path 1 to reach intrinsic/ecological antifragility}: \textbf{Mathematical identification of second-order effects} in the system response characterizing antifragile behaviour;
    \item \textit{Path 2 to reach inherited/evolutionary antifragility}: \textbf{Mapping the dynamics of the system to physical principles of criticality and evolution} to describe reaching an antifragile state;
    \item \textit{Path 3 to reach induced/interventional antifragility}: \textbf{Nonlinear control synthesis or learning of optimal driving signals} to push dynamical systems to antifragile regions in their response spectrum;
\end{itemize}

It is important to note that, \textit{technical systems are typically built as homogeneous systems}, i.e., all components share the same properties (i.e., spatial, temporal, structural, and functional). \textit{However, most natural systems are heterogeneous, where a few components are closer, faster, stronger, or more relevant than others}. Even if we analyze across scales, the common denominator is time. Each scale or property of the system evolves, be it slower or faster. When defining antifragility, time plays an important role. The study of natural systems within the antifragility concept needs special treatment as we typically use limited models for complex, nonlinear natural phenomena and processes where the design principles are rooted in evolution, as described in the work on resilience, reactivity, and variability of Arnoldi et al. \cite{arnoldi2016resilience}, the control theoretic perspective of Marom et al.\cite{marom2023biophysical}, and the geometric robustness of Ay et al.\cite{ay2007geometric}. The study of technical systems differs because we better understand their design principles. Therefore, \textit{before discussing whether a system is antifragile or not, we need to define the environment in which the system operates and the objectives of the system}. Finally, it should be noted that given the environment and the objectives, there may be multiple payoff functions that can be considered for quantification of the system's performance, as shown in the systematic work of Kraskova et al.\cite{krakovska2021resilience}. Depending on the indicators or metrics we adopt, the conclusions regarding the system's fragility can differ. Consequently, in the following section, we will discuss the differences between intrinsic, inherited, and induced antifragility, given certain specifications and intuitive examples in both technical and natural systems.

\section{Intrinsic and ecological antifragility}
Intrinsic/ecological antifragility (or fragility) quantifies the benefit (or harm) of input distribution unevenness, volatility, or perturbations attributed to the nonlinearity of the system's payoff function.

\begin{figure}[!b]
    \centering
    \includegraphics[width=0.5\linewidth]{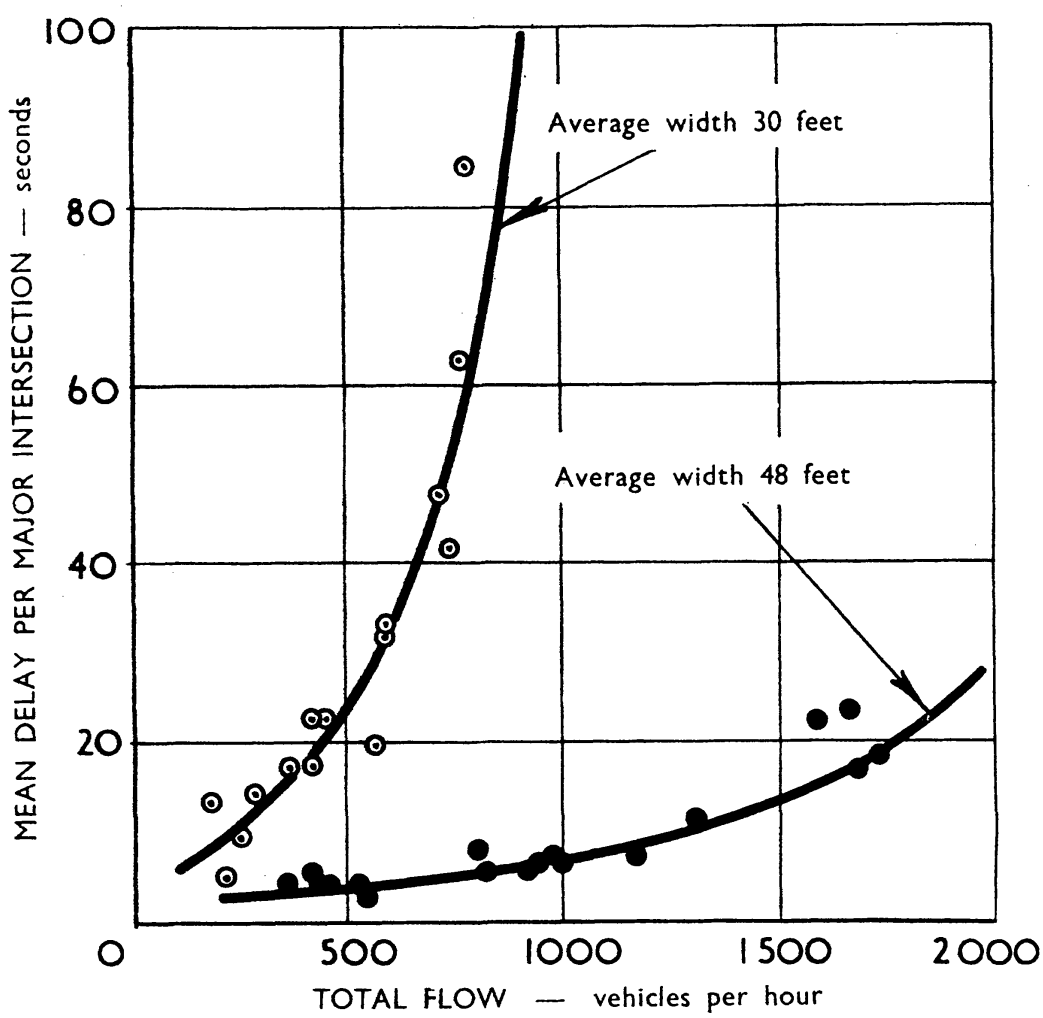}
    \caption{Source: Wardrop \cite{wardrop1954capacity} - Delay at a major intersection of two groups of streets in Central London.}
    \label{fig:delays}
\end{figure}

\subsection{Technical systems: intrinsic (anti)-fragility}
When considering intrinsic antifragility, the fundamental aspect is dynamics' time-scale separation describing the input-output coupling. An example is shown in figure~\ref{fig:delays} which illustrates the fragile operation of the transport systems, where exponentially increasing delays (output) occur with the increasing flow (input)\cite{wardrop1954capacity}. Management strategies, such as those suggested in \cite{papageorgiou2003review}, aim to push the operation of the system towards the critical density. Such strategies consider the effect of variation in input (traffic flow) on the outcome (delays), which are modulated by the shape of the input-output function (exponential), as illustrated in schematic figure \ref{fig:spectrum} schematic).

In another example, given the interactions between multiple control loops (e.g., the internal DC motor control loop and outer robot position control loop\cite{axenie2023antifragilerobotics}, the oscillators models of traffic flow inputs\cite{axenie2023antifragiletraffic}, and the coupled traffic dynamics \cite{sun2023exploring,makridis2023rule}), the definition and quantification of antifragility emerge in the context of input-output mappings of time scales separation performed to handle uncertainty and high-frequency phenomena. 
However, please note that such systems employ intrinsic antifragility to achieve their prescribed objectives by design, given the specifications and constraints imposed by the physics of the operational space of the system, as described amply in the work on driver models of Kesting et al.\cite{kesting2010enhanced, milanes2014modeling, makridis2019mfc} and traffic dynamics, as in the work of Du et al.\cite{du2023adaptive}, Daganzo et al.\cite{daganzo1994cell}, and Messner et al.\cite{messner1990metanet}. 

When unpredictable fluctuations start developing in a dynamical system that is on the verge of losing its stability, it is commonly referred to as criticality. Intrinsic antifragility can be cast around a system's criticality. The ability to absorb and respond to stresses, due to the appearance of scale-free temporal fluctuations, slowing dynamics, and multistability, are the fundamental indicators of criticality. Here, antifragility can be viewed as the motion of a system from an existing steady state to a better one in the aftermath of a change in conditions. This transfer among steady states may be continuous (second-order phase transition) or discontinuous (first-order phase transition), as formalized in the work of Schaffer et al.\cite{scheffer2009early}.

\subsection{Natural systems: ecological (anti)-fragility}
\begin{figure}[t]
    \centering
    \includegraphics[width=0.9\linewidth]{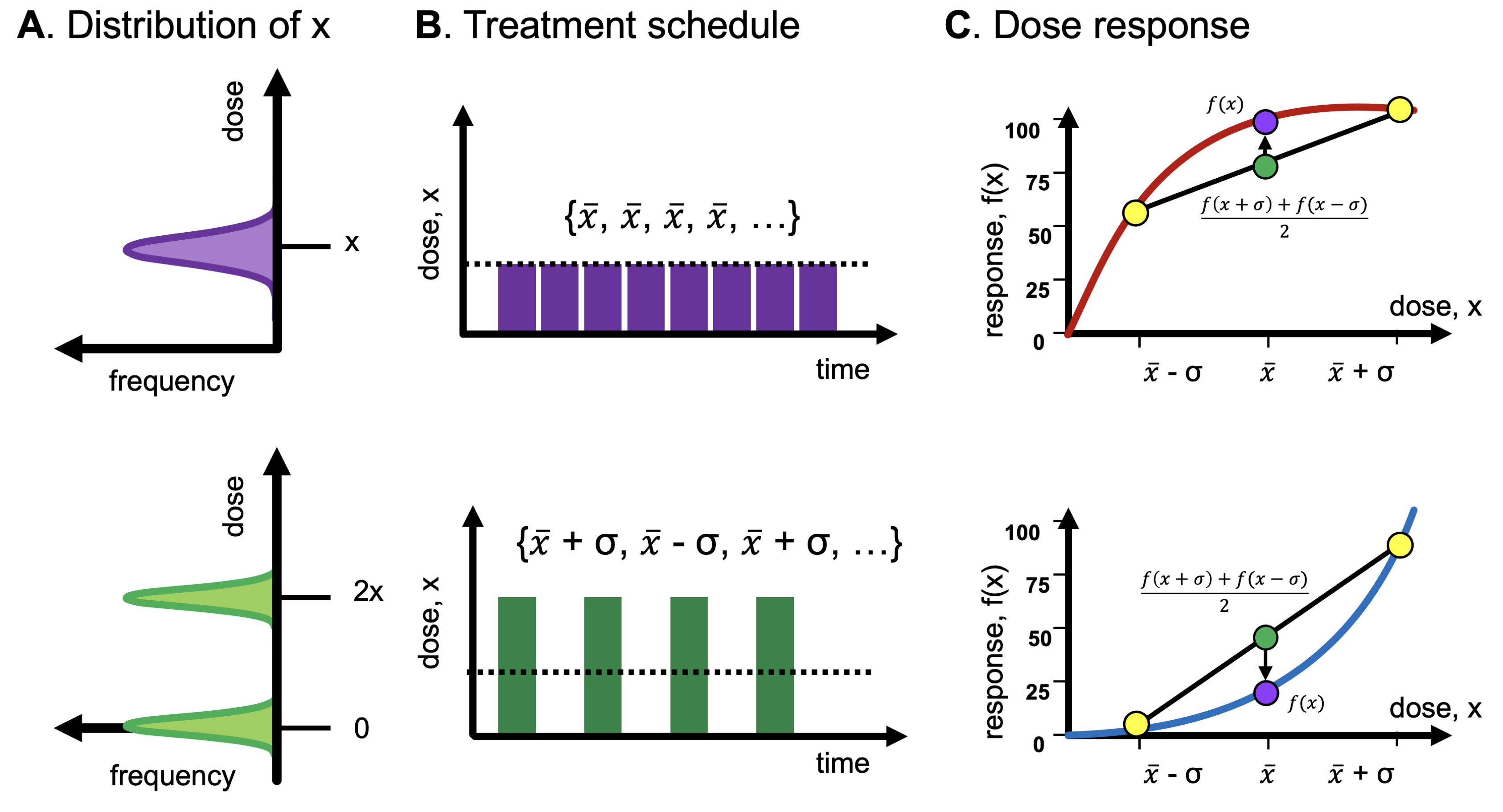}
    \caption{Figure reproduced from Taleb, 2023 \cite{taleb2023working} - Example treatment-scheduling protocols. (A) Example dose distribution with low  (top) or high variance (bottom). (B) Associated protocols. (C) Low variance protocols are optimal to maximize response for concavity; high variance protocols for convexity.}
    \label{fig:taleb}
\end{figure}

In biological systems, changes in environmental conditions can decrease the survival and fecundity of individuals based on a species' Darwinian fitness. The rate of change in environmental perturbations may reduce fitness in response to stochastic fluctuations and seasonal variation, as formalized in the work of Taleb\cite{taleb2018anti}. The payoff function associated with the system response to environmental variation may be concave (fragile), convex (antifragile), or linear (neutral), as suggested in the work of Danchin et al.\cite{danchin2011antifragility} and Taleb \cite{taleb2012antifragile}. An example is shown in figure \ref{fig:taleb}, where there exists a dose distribution mean and variance (panel A) associated with each anti-cancer treatment protocol (panel B), where the outcome depends on the concave (top) or convex (bottom) dose-response function (panel C). Concave functions (red) should employ low-variance protocols, while concave functions (blue) should employ high-variance intermittent protocols. Ecological antifragility is the natural systems correlate to intrinsic antifragility and may defined as the system benefit derived from input volatility\cite{taleb2023working}. Defining ecological fragility or antifragility is useful for the control of a biological population. For example, mathematical models describing the response to anti-cancer drugs measure the ecological effect of volatile versus continuous treatment schedules, as shown in the work of West et al.\cite{west2020antifragile} or the pharmacokinetics study of Pierik et al.\cite{pierik2023second}. 

In this context one can also build a rigorous description of all aspects of ecosystem antifragility and its applications, as detailed in the work of Equihua et al. \cite{equihua2020ecosystem}. Antifragility requires optimal computational and inference capabilities that most interestingly occur within another fundamental concept of complex systems: criticality. As highlighted by Hidalgo \cite{hidalgo2014information}, empirical investigations have suggested that living systems operate in the proximity of critical thresholds, existing at the delicate boundary between order and randomness demonstrated across various domains including electrical heart activity and brain function, among others \cite{kiyono2005phase,ivanov1996scaling,rivera2016heart,goldberger2002physiologic}. Precise measurement of the payoff function for predicting antifragility plays a key role. L\'opez-Corona and coworkers\cite{lopez2022esd} applied these ideas discussed above to the scale of planetary ecosystem antifragility by integrating well-established principles from non-equilibrium thermodynamics and adopting a system dynamics approach using Fisher's information on Earth's entropy production\cite{Fernandez2014,lopez2018forest}.

\section{Inherited \& evolutionary antifragility}
Inherited/evolutionary antifragility (or fragility) quantifies the benefit (or harm) of input distribution unevenness, volatility, or perturbations attributed to the system's response to external signals.

\subsection{Technical systems: inherited (anti)-fragility}
Antifragility is correlated with increased system heterogeneity at the inherited scale, as shown in the work of Lopez-Otrasenda et al.\cite{lopez2023temporal} and Makridis et al.\cite{makridis2020formalizing}. This aspect comes into play when considering the system's capacity to build extra capacity in anticipation of perturbations. More precisely, to achieve inherited antifragility, a system designer builds upon the time-scale separation of a redundant overcompensation component, as shown in robotics\cite{axenie2023antifragilerobotics}, traffic control\cite{axenie2023antifragiletraffic}, and medical\cite{axenie2022antifragileoncology} instantiations of applied antifragility. 

Similarly, machine learning systems developed in the work of Sun et al.\cite{sun2023exploring} maintain time scale separation through the formulation of the learning task. More precisely, the system determines the best action at which the system yields the maximum possible discounted future reward. Isolating only the contribution of the time scale separation and the redundant overcompensation term, the machine learning inherited antifragility controller gained in the skewness (i.e., convexity) of the disruption magnitude over surging demand. In such systems, antifragility is quantified as the geometric properties of the anticipating control actions and the shape of the response in the face of high-magnitude external perturbations.

Finally, criticality also plays a role in inherited antifragility by allowing a system to leave the current steady state, for a different one. The trigger for state switching may come from a change in the parameters of the system, noise being externally enforced, or from an increase in the rates of the system from neighbouring entities in a competition of cooperation with the system, as shown in the works of Angeli et al.\cite{angeli2004detection} and Hizanidis et al.\cite{hizanidis2008delay}.

\subsection{Natural systems: evolutionary (anti)-fragility}
Evolution, defined here as the change in heritable traits within a population over time, is also influenced by environmental perturbations\cite{lopez2019rise}. In the previous section, ecological antifragility considered individual species in isolation to quantify the response to perturbation. Evolutionary antifragility quantifies how a population of interacting species is affected by perturbations. For example, in cancer, competition between heterogeneous populations of cell types modulates antifragility \cite{bayer2023games}. In natural biological systems, we investigate complexity (which implies maximum computational capabilities) and how systems reach criticality \cite{crosato2018critical, gershenson2012complexity, kalloniatis2018fisher}. Adaptive mechanisms of living systems do more than merely react to the environment's variability through random mutations followed by selection; they must have built-in characteristics that enable them to discover alternatives to cope with adversity, variability, and uncertainty\cite{equihua2020ecosystem}. 

Using theoretical arguments, it has been proposed that systems under eco-evolution tend to be at criticality, implying maximum complexity and computational and inferential capabilities; and then they are also at maximum antifragility \cite{equihua2020ecosystem,lopez2019fisher,pineda2019novel}. Stability plays a central role for function, in both natural and technical dynamical systems. Paradoxically, the only way a dynamical system can remain stable is if it is excitable and able to change its behavior in reaction to outside stimuli. It is flexible, thus it is stable; in fact, the organism's true stability depends on its modest instability, as beautifully described by Cannon\cite{cannon1929organization}. 

Inherited antifragility in natural systems is a consequence of the self-organized configuration of a natural system that arises from the interactions among all components through evolutionary processes (e.g., genetic inheritance, microbiome inheritance, and social inheritance), constrained by external conditions. Antifragile natural systems derive benefits under uncertainty, stressors, and perturbations in both ecological and evolutionary scales. Evolution by natural selection itself can be thought of as an antifragile process, whereby a population is maintained amidst environmental perturbations through genetic variation. For example, although individuals within a population may die, the population evolves toward a more antifragile state, increasingly higher fitness to address fluctuating environmental conditions\cite{danchin2011antifragility, nichol2016stochasticity}.

\section{Induced \& interventional antifragility}
Induced/interventional antifragility (or fragility) quantifies the benefit (or harm) of input distribution unevenness, volatility, or perturbations attributed to the system's response to closed-loop controllers.

\subsection{Technical systems: induced (anti)-fragility}
\begin{figure}[t]
    \centering
    \includegraphics[width=0.9\linewidth]{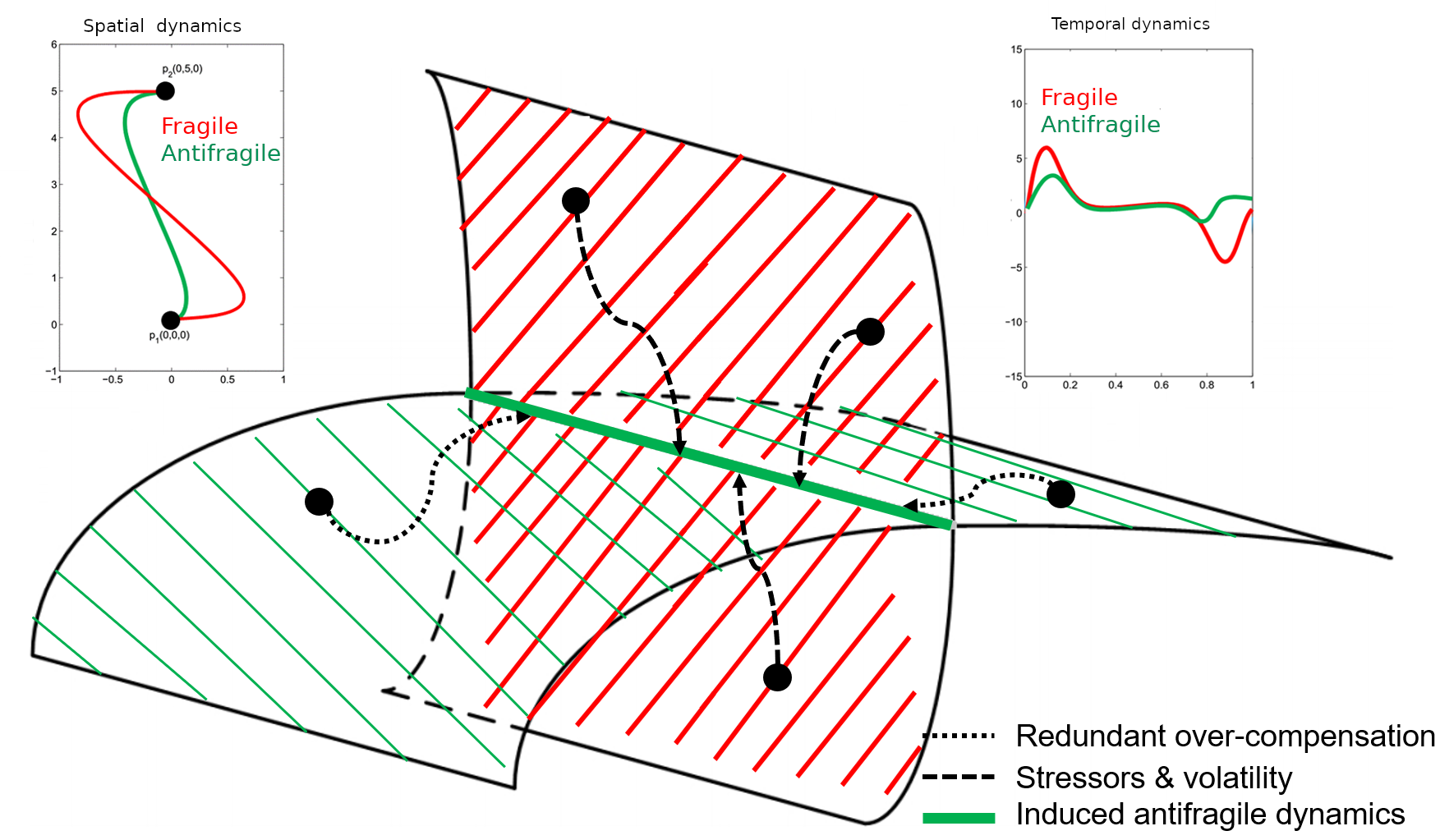}
    \caption{Figure reproduced from Axenie et Saveriano\cite{axenie2023antifragilerobotics} - Spatial and temporal dynamics of fragile and antifragile behaviours of a robot in uncertain environments. Closed-loop dynamics in the presence of stressors and volatility.}
    \label{fig:slidingmode}
\end{figure}

The control theoretic approach to induced antifragility focuses on the combination of time-scale separation and redundant overcompensation with variable structure control. Put simply, the system is now pushed to the antifragile region of its operational domain through a judicious choice of an external control or regulation signal, as shown by Axenie and Saveriano\cite{axenie2023antifragilerobotics}. This can be accomplished by properly synthesizing a control law, which develops a redundant over-compensation capacity to handle uncertainty regarding the sensor and actuator failures by pushing the closed-loop system dynamics to prescribed dynamics. This is captured in Figure~\ref{fig:slidingmode}, where the fragile--antifragile behavior is depicted across spatiotemporal dynamics of a robot in uncertain environments. 

Antifragility was quantified in this case through the quality of the dynamics tracking and the speed of reaching the desired region of the desired dynamics manifold in the presence of uncertainty and volatility through adaptive control\cite{wang2019path}, robust control\cite{solea2007trajectory}, and resilient control\cite{antonelli2007fuzzy} strategies. In a large-scale traffic control instantiation by Axenie et al.\cite{axenie2023antifragiletraffic}, the antifragile controller demonstrated statistically significant gains given increasing traffic disruptions amplitude over time. The systematic evaluation demonstrates that the control law selection based on the second-order effects of the signal re-computation may capture the volatile dynamics of the closed-loop system. 

Considering machine learning approaches to induced antifragility, the work of Sun et al.\cite{sun2023exploring} designed a traffic reinforcement learning agent that learns to be conservative when regulating the controlled region. Here, we have again a clear quantification of the system's antifragility based on a dynamics response curve to external uncertainty (i.e., amplitude of traffic disruptions) and volatility (i.e., onset and offset of traffic disruptions) overcoming the baseline approach (i.e., static police-made traffic light control), a state-of-the-art model predictive control\cite{geroliminis2012optimal}, and other reinforcement learning approaches\cite{zhou2023scalable}.

Finally, when considering criticality, the control of multi-stability deals with the transition of the system to a more desired steady state and also preventing it from moving to an inferior one, as shown in the work of Pisarchik et al.\cite{pisarchik2014control}. From the control point of view, it is beneficial to predict the tipping points well before they occur so that remedial actions can be adopted, as demonstrated in the work of Grziwotz et al.\cite{grziwotz2023anticipating}.

\subsection{Natural systems: interventional (anti)-fragility}
Eradication of a heterogeneous population may be exceedingly difficult due to the evolution of resistance to interventional treatments. For example, the continuous administration of anti-cancer drugs\cite{gatenby2009adaptive, read2011evolution} or antibiotics\cite{read2014antibiotic} selects for resistant sub-populations rendering subsequent treatments ineffective. Recent work to apply principles learned from agricultural methods known as ``Integrative Pest Management'' (IPM) has shown some success in the management of cancer\cite{whelan2020resistance, cunningham2019call}. For example, adaptive cancer therapy uses a simple rule-of-thumb protocol to adapt treatment administration and treatment break time intervals based on tumor response\cite{zhang2017integrating}. Importantly, this adaptive protocol in an increase in dose variance (prolonged periods of high dose followed by prolonged periods of zero dose)\cite{west2023fundamentals}. This is part of a broader effort to design treatment protocols using evolutionary principles that increase the treatment-induced volatility that tumor cells undergo during the course of treatment to maximize tumor regression\cite{strobl2023treatment}. Mathematical models of tumor-immune-drug interactions can drive chemotherapy optimization regimens to maximize the efficacy/toxicity ratio, as shown by West et al.\cite{west2019multidrug} and Axenie et al.\cite{axenie2022antifragileoncology}. 

In another interventional example, in ecological processes of terrestrial ecosystems, the complexity of interactions, and costs involved, among other technical complications, but also, maybe more importantly, due to ethical concerns because of path dependence, computational irreducibility, and unseen consequences, there are very few interventions that aim to restore ecosystem antifragility, as shown in the work of Ripple et al.\cite{ripple2012trophic}, Beschta et al.\cite{beschta2016riparian}, and Wright et al.\cite{wright2002ecosystem}.

Furthermore, a series of recent papers\cite{ramirez2023similar, isaac2023potential} have shown that dietary patterns might influence network communication along the brain-gut axis, especially at the age that both systems go through maturation processes. From an ecological perspective, an adequate level of connectivity dissipates the effect of perturbations in the distribution of species and enhances ecosystem stability. A loss in connectivity leads to a loss in gut microbiota ecosystem antifragility. The basic rationale is that a system's response to perturbation requires an efficient flow of information. For maximum antifragility, this flow must be optimal, implying maximum connectivity, as shown in the papers of Isaac et al.\cite{isaac2023potential} and Equihua et al.\cite{equihua2020ecosystem}.

\section{Discussion}
In this perspective, we have reviewed efforts to apply antifragility theory to both technical/physical systems as well as natural/biological systems. We provide a conceptual framework to unify the language across both systems and define the relevant scales of fragility, summarized below.

{\bf Trademarks of intrinsic \& ecological antifragility}: Intrinsic and ecological antifragile systems benefit from harm derived from internal dynamics distribution unevenness, based on the convexity of the response function of the system without external input and solely based on the internal components' heterogeneity and resilience. Features such as stability describe the most simple system response with minimal antifragile characteristics. Within this scale, precise characterization of the payoff function describing the relationship between system inputs and outputs is of critical importance. 

{\bf Trademarks of inherited \& evolutionary antifragility}: Homogeneity and heterogeneity play a crucial role in the design and synthesis of inherited and evolutionary antifragile systems. From criticality and multi-level interactions of multiple time scales to quantifying criticality margins, such a design scheme leverages local interactions of the system to make it benefit from perturbations. In other words, inherited and evolutionary antifragile systems benefit from harm derived from input distribution unevenness, based on the emergent system dynamics and its interactions with the operating environment (i.e. disturbances, noise, modulated perturbations).

{\bf Trademarks of induced \& interventional antifragility}: Inducing a desired behaviour within interventional antifragility requires an innovative, control-theoretic, design and synthesis approach. Here, nonlinear dynamics across both space and time can promote the system's capacity to absorb internal and external disruptions. Induced and interventional antifragile systems benefit from harm derived from input distribution unevenness based on emergent system dynamics in closed-loop with a controller driving the system towards prescribed dynamics in the presence of modulated or non-stationary disturbances, noise, and volatility.

It is also important to note that we need not constrain ourselves to search only for systems that are antifragile. First, a system may be fragile on the intrinsic or inherited scales yet still be amenable to interventions that are antifragile through clever design of feedback controllers. Second, fragility (or antifragility) is a measurable quantity: the response of a system to volatility. Some systems (e.g., tumors) may respond in a fragile manner to given perturbations (e.g., chemotherapy) and thus it is critical to characterize the system along the fragile-antifragile spectrum. Finally, though we consider a discrete layering of antifragility types, we are aware that this structure can be applied systematically across scales. In other words, even the simplest systems can be subject to all three types of antifragility.

\section{Acknowledgments}
Meisam Akbarzadeh has been supported by ETH Zurich Risk Center. Jeffrey West is supported by funding from the National Cancer Institute via the Cancer Systems Biology Consortium (CSBC) U54CA274507 and from the Moffitt Center of Excellence for Evolutionary Therapy.

\newpage
\bibliography{bibliography.bib}

\begin{thebibliography}{10}
\urlstyle{rm}
\expandafter\ifx\csname url\endcsname\relax
  \def\url#1{\texttt{#1}}\fi
\expandafter\ifx\csname urlprefix\endcsname\relax\def\urlprefix{URL }\fi
\expandafter\ifx\csname doiprefix\endcsname\relax\def\doiprefix{DOI: }\fi
\providecommand{\bibinfo}[2]{#2}
\providecommand{\eprint}[2][]{\url{#2}}

\bibitem{taleb2012antifragile}
\bibinfo{author}{Taleb, N.~N.}
\newblock \emph{\bibinfo{title}{Antifragile: Things that gain from disorder}}, vol.~\bibinfo{volume}{3} (\bibinfo{publisher}{Random House Incorporated}, \bibinfo{year}{2012}).

\bibitem{taleb2013mathematical}
\bibinfo{author}{Taleb, N.~N.} \& \bibinfo{author}{Douady, R.}
\newblock \bibinfo{journal}{\bibinfo{title}{Mathematical definition, mapping, and detection of (anti) fragility}}.
\newblock {\emph{\JournalTitle{Quantitative Finance}}} \textbf{\bibinfo{volume}{13}}, \bibinfo{pages}{1677--1689} (\bibinfo{year}{2013}).

\bibitem{axenie2023antifragilerobotics}
\bibinfo{author}{Axenie, C.} \& \bibinfo{author}{Saveriano, M.}
\newblock \bibinfo{journal}{\bibinfo{title}{Antifragile control systems: The case of mobile robot trajectory tracking in the presence of uncertainty}}.
\newblock {\emph{\JournalTitle{arXiv preprint arXiv:2302.05117}}}  (\bibinfo{year}{2023}).

\bibitem{pineda2018Antifragility}
\bibinfo{author}{Pineda, O.~K.}, \bibinfo{author}{Kim, H.} \& \bibinfo{author}{Gershenson, C.}
\newblock \bibinfo{journal}{\bibinfo{title}{Antifragility of random boolean networks}}.
\newblock {\emph{\JournalTitle{arXiv preprint arXiv:1812.06760}}}  (\bibinfo{year}{2018}).

\bibitem{de2020antifragility}
\bibinfo{author}{de~Bruijn, H.}, \bibinfo{author}{Groessler, A.} \& \bibinfo{author}{Videira, N.}
\newblock \bibinfo{journal}{\bibinfo{title}{Antifragility as a design criterion for modelling dynamic systems}}.
\newblock {\emph{\JournalTitle{Systems Research and Behavioral Science}}} \textbf{\bibinfo{volume}{37}}, \bibinfo{pages}{23--37} (\bibinfo{year}{2020}).

\bibitem{blevcic2020antifragile}
\bibinfo{author}{Ble{\v{c}}i{\'c}, I.} \& \bibinfo{author}{Cecchini, A.}
\newblock \bibinfo{journal}{\bibinfo{title}{Antifragile planning}}.
\newblock {\emph{\JournalTitle{Planning Theory}}} \textbf{\bibinfo{volume}{19}}, \bibinfo{pages}{172--192} (\bibinfo{year}{2020}).

\bibitem{johnson2013antifragility}
\bibinfo{author}{Johnson, J.} \& \bibinfo{author}{Gheorghe, A.~V.}
\newblock \bibinfo{journal}{\bibinfo{title}{Antifragility analysis and measurement framework for systems of systems}}.
\newblock {\emph{\JournalTitle{International Journal of Disaster Risk Science}}} \textbf{\bibinfo{volume}{4}}, \bibinfo{pages}{159--168} (\bibinfo{year}{2013}).

\bibitem{arnoldi2016resilience}
\bibinfo{author}{Arnoldi, J.-F.}, \bibinfo{author}{Loreau, M.} \& \bibinfo{author}{Haegeman, B.}
\newblock \bibinfo{journal}{\bibinfo{title}{Resilience, reactivity and variability: A mathematical comparison of ecological stability measures}}.
\newblock {\emph{\JournalTitle{Journal of Theoretical Biology}}} \textbf{\bibinfo{volume}{389}}, \bibinfo{pages}{47--59} (\bibinfo{year}{2016}).

\bibitem{marom2023biophysical}
\bibinfo{author}{Marom, S.} \& \bibinfo{author}{Marder, E.}
\newblock \bibinfo{journal}{\bibinfo{title}{A biophysical perspective on the resilience of neuronal excitability across timescales}}.
\newblock {\emph{\JournalTitle{Nature Reviews Neuroscience}}} \textbf{\bibinfo{volume}{24}}, \bibinfo{pages}{640--652} (\bibinfo{year}{2023}).

\bibitem{ay2007geometric}
\bibinfo{author}{Ay, N.} \& \bibinfo{author}{Krakauer, D.~C.}
\newblock \bibinfo{journal}{\bibinfo{title}{Geometric robustness theory and biological networks}}.
\newblock {\emph{\JournalTitle{Theory in Biosciences}}} \textbf{\bibinfo{volume}{125}}, \bibinfo{pages}{93--121} (\bibinfo{year}{2007}).

\bibitem{krakovska2021resilience}
\bibinfo{author}{Krakovsk{\'a}, H.}, \bibinfo{author}{Kuehn, C.} \& \bibinfo{author}{Longo, I.~P.}
\newblock \bibinfo{journal}{\bibinfo{title}{Resilience of dynamical systems}}.
\newblock {\emph{\JournalTitle{European Journal of Applied Mathematics}}} \bibinfo{pages}{1--46} (\bibinfo{year}{2021}).

\bibitem{wardrop1954capacity}
\bibinfo{author}{Wardrop, J.~G.}
\newblock \bibinfo{journal}{\bibinfo{title}{The capacity of roads}}.
\newblock {\emph{\JournalTitle{Journal of the Operational Research Society}}} \textbf{\bibinfo{volume}{5}}, \bibinfo{pages}{14--24} (\bibinfo{year}{1954}).

\bibitem{papageorgiou2003review}
\bibinfo{author}{Papageorgiou, M.}, \bibinfo{author}{Diakaki, C.}, \bibinfo{author}{Dinopoulou, V.}, \bibinfo{author}{Kotsialos, A.} \& \bibinfo{author}{Wang, Y.}
\newblock \bibinfo{journal}{\bibinfo{title}{Review of road traffic control strategies}}.
\newblock {\emph{\JournalTitle{Proceedings of the IEEE}}} \textbf{\bibinfo{volume}{91}}, \bibinfo{pages}{2043--2067} (\bibinfo{year}{2003}).

\bibitem{axenie2023antifragiletraffic}
\bibinfo{author}{Axenie, C.} \& \bibinfo{author}{Grossi, M.}
\newblock \bibinfo{journal}{\bibinfo{title}{Antifragile control systems: The case of an oscillator-based network model of urban road traffic dynamics}}.
\newblock {\emph{\JournalTitle{arXiv preprint arXiv:2210.10460}}}  (\bibinfo{year}{2023}).

\bibitem{sun2023exploring}
\bibinfo{author}{Sun, L.} \emph{et~al.}
\newblock \bibinfo{title}{Exploring antifragility in traffic networks: Anticipating disturbances with reinforcement learning}.
\newblock In \emph{\bibinfo{booktitle}{23rd Swiss Transport Research Conference (STRC 2023)}} (\bibinfo{year}{2023}).

\bibitem{makridis2023rule}
\bibinfo{author}{Makridis, M.~A.}, \bibinfo{author}{Schaniel, J.} \& \bibinfo{author}{Kouvelas, A.}
\newblock \bibinfo{journal}{\bibinfo{title}{Rule-based on-off traffic control strategy for cavs on motorway networks: Assessing cooperation level and driving homogeneity}}.
\newblock {\emph{\JournalTitle{IEEE Access}}}  (\bibinfo{year}{2023}).

\bibitem{kesting2010enhanced}
\bibinfo{author}{Kesting, A.}, \bibinfo{author}{Treiber, M.} \& \bibinfo{author}{Helbing, D.}
\newblock \bibinfo{journal}{\bibinfo{title}{Enhanced intelligent driver model to access the impact of driving strategies on traffic capacity}}.
\newblock {\emph{\JournalTitle{Philosophical Transactions of the Royal Society A: Mathematical, Physical and Engineering Sciences}}} \textbf{\bibinfo{volume}{368}}, \bibinfo{pages}{4585--4605} (\bibinfo{year}{2010}).

\bibitem{milanes2014modeling}
\bibinfo{author}{Milan{\'e}s, V.} \& \bibinfo{author}{Shladover, S.~E.}
\newblock \bibinfo{journal}{\bibinfo{title}{Modeling cooperative and autonomous adaptive cruise control dynamic responses using experimental data}}.
\newblock {\emph{\JournalTitle{Transportation Research Part C: Emerging Technologies}}} \textbf{\bibinfo{volume}{48}}, \bibinfo{pages}{285--300} (\bibinfo{year}{2014}).

\bibitem{makridis2019mfc}
\bibinfo{author}{Makridis, M.}, \bibinfo{author}{Fontaras, G.}, \bibinfo{author}{Ciuffo, B.} \& \bibinfo{author}{Mattas, K.}
\newblock \bibinfo{journal}{\bibinfo{title}{Mfc free-flow model: Introducing vehicle dynamics in microsimulation}}.
\newblock {\emph{\JournalTitle{Transportation Research Record}}} \textbf{\bibinfo{volume}{2673}}, \bibinfo{pages}{762--777} (\bibinfo{year}{2019}).

\bibitem{du2023adaptive}
\bibinfo{author}{Du, Y.}, \bibinfo{author}{Makridis, M.~A.}, \bibinfo{author}{Tamp{\`e}re, C.~M.}, \bibinfo{author}{Kouvelas, A.} \& \bibinfo{author}{ShangGuan, W.}
\newblock \bibinfo{journal}{\bibinfo{title}{Adaptive control with moving actuators at motorway bottlenecks with connected and automated vehicles}}.
\newblock {\emph{\JournalTitle{Transportation Research Part C: Emerging Technologies}}} \textbf{\bibinfo{volume}{156}}, \bibinfo{pages}{104319} (\bibinfo{year}{2023}).

\bibitem{daganzo1994cell}
\bibinfo{author}{Daganzo, C.~F.}
\newblock \bibinfo{journal}{\bibinfo{title}{The cell transmission model: A dynamic representation of highway traffic consistent with the hydrodynamic theory}}.
\newblock {\emph{\JournalTitle{Transportation Research Part B: Methodological}}} \textbf{\bibinfo{volume}{28}}, \bibinfo{pages}{269--287} (\bibinfo{year}{1994}).

\bibitem{messner1990metanet}
\bibinfo{author}{Messner, A.} \& \bibinfo{author}{Papageorgiou, M.}
\newblock \bibinfo{journal}{\bibinfo{title}{Metanet: A macroscopic simulation program for motorway networks}}.
\newblock {\emph{\JournalTitle{Traffic Engineering \& Control}}} \textbf{\bibinfo{volume}{31}}, \bibinfo{pages}{466--470} (\bibinfo{year}{1990}).

\bibitem{scheffer2009early}
\bibinfo{author}{Scheffer, M.} \emph{et~al.}
\newblock \bibinfo{journal}{\bibinfo{title}{Early-warning signals for critical transitions}}.
\newblock {\emph{\JournalTitle{Nature}}} \textbf{\bibinfo{volume}{461}}, \bibinfo{pages}{53--59} (\bibinfo{year}{2009}).

\bibitem{taleb2018anti}
\bibinfo{author}{Taleb, N.~N.}
\newblock \bibinfo{title}{(anti) fragility and convex responses in medicine}.
\newblock In \emph{\bibinfo{booktitle}{International Conference on Complex Systems}}, \bibinfo{pages}{299--325} (\bibinfo{organization}{Springer}, \bibinfo{year}{2018}).

\bibitem{danchin2011antifragility}
\bibinfo{author}{Danchin, A.}, \bibinfo{author}{Binder, P.~M.} \& \bibinfo{author}{Noria, S.}
\newblock \bibinfo{journal}{\bibinfo{title}{Antifragility and tinkering in biology (and in business) flexibility provides an efficient epigenetic way to manage risk}}.
\newblock {\emph{\JournalTitle{Genes}}} \textbf{\bibinfo{volume}{2}}, \bibinfo{pages}{998--1016} (\bibinfo{year}{2011}).

\bibitem{taleb2023working}
\bibinfo{author}{Taleb, N.~N.} \& \bibinfo{author}{West, J.}
\newblock \bibinfo{journal}{\bibinfo{title}{Working with convex responses: Antifragility from finance to oncology}}.
\newblock {\emph{\JournalTitle{Entropy}}} \textbf{\bibinfo{volume}{25}}, \bibinfo{pages}{343} (\bibinfo{year}{2023}).

\bibitem{west2020antifragile}
\bibinfo{author}{West, J.} \emph{et~al.}
\newblock \bibinfo{journal}{\bibinfo{title}{Antifragile therapy}}.
\newblock {\emph{\JournalTitle{BioRxiv}}} \bibinfo{pages}{2020--10} (\bibinfo{year}{2020}).

\bibitem{pierik2023second}
\bibinfo{author}{Pierik, L.}, \bibinfo{author}{McDonald, P.}, \bibinfo{author}{Anderson, A.~R.} \& \bibinfo{author}{West, J.}
\newblock \bibinfo{journal}{\bibinfo{title}{Second-order effects of chemotherapy pharmacodynamics and pharmacokinetics on tumor regression and cachexia}}.
\newblock {\emph{\JournalTitle{bioRxiv}}} \bibinfo{pages}{2023--06} (\bibinfo{year}{2023}).

\bibitem{equihua2020ecosystem}
\bibinfo{author}{Equihua, M.} \emph{et~al.}
\newblock \bibinfo{journal}{\bibinfo{title}{Ecosystem antifragility: beyond integrity and resilience}}.
\newblock {\emph{\JournalTitle{PeerJ}}} \textbf{\bibinfo{volume}{8}}, \bibinfo{pages}{e8533} (\bibinfo{year}{2020}).

\bibitem{hidalgo2014information}
\bibinfo{author}{Hidalgo, J.} \emph{et~al.}
\newblock \bibinfo{journal}{\bibinfo{title}{Information-based fitness and the emergence of criticality in living systems}}.
\newblock {\emph{\JournalTitle{Proceedings of the National Academy of Sciences}}} \textbf{\bibinfo{volume}{111}}, \bibinfo{pages}{10095--10100} (\bibinfo{year}{2014}).

\bibitem{kiyono2005phase}
\bibinfo{author}{Kiyono, K.}, \bibinfo{author}{Struzik, Z.~R.}, \bibinfo{author}{Aoyagi, N.}, \bibinfo{author}{Togo, F.} \& \bibinfo{author}{Yamamoto, Y.}
\newblock \bibinfo{journal}{\bibinfo{title}{Phase transition in a healthy human heart rate}}.
\newblock {\emph{\JournalTitle{Physical Review Letters}}} \textbf{\bibinfo{volume}{95}}, \bibinfo{pages}{058101} (\bibinfo{year}{2005}).

\bibitem{ivanov1996scaling}
\bibinfo{author}{Ivanov, P.~C.} \emph{et~al.}
\newblock \bibinfo{journal}{\bibinfo{title}{Scaling behaviour of heartbeat intervals obtained by wavelet-based time-series analysis}}.
\newblock {\emph{\JournalTitle{Nature}}} \textbf{\bibinfo{volume}{383}}, \bibinfo{pages}{323--327} (\bibinfo{year}{1996}).

\bibitem{rivera2016heart}
\bibinfo{author}{Rivera, A.~L.} \emph{et~al.}
\newblock \bibinfo{journal}{\bibinfo{title}{Heart rate and systolic blood pressure variability in the time domain in patients with recent and long-standing diabetes mellitus}}.
\newblock {\emph{\JournalTitle{PloS One}}} \textbf{\bibinfo{volume}{11}}, \bibinfo{pages}{e0148378} (\bibinfo{year}{2016}).

\bibitem{goldberger2002physiologic}
\bibinfo{author}{Goldberger, A.~L.}, \bibinfo{author}{Peng, C.-K.} \& \bibinfo{author}{Lipsitz, L.~A.}
\newblock \bibinfo{journal}{\bibinfo{title}{What is physiologic complexity and how does it change with aging and disease?}}
\newblock {\emph{\JournalTitle{Neurobiology of Aging}}} \textbf{\bibinfo{volume}{23}}, \bibinfo{pages}{23--26} (\bibinfo{year}{2002}).

\bibitem{lopez2022esd}
\bibinfo{author}{L{\'o}pez-Corona, O.}, \bibinfo{author}{Kolb, M.}, \bibinfo{author}{Ram{\'\i}rez-Carrillo, E.} \& \bibinfo{author}{Lovett, J.}
\newblock \bibinfo{journal}{\bibinfo{title}{Esd ideas: planetary antifragility: a new dimension in the definition of the safe operating space for humanity}}.
\newblock {\emph{\JournalTitle{Earth System Dynamics}}} \textbf{\bibinfo{volume}{13}}, \bibinfo{pages}{1145--1155} (\bibinfo{year}{2022}).

\bibitem{Fernandez2014}
\bibinfo{author}{Fern{\'a}ndez, N.} \& \bibinfo{author}{Gershenson, C.}
\newblock \bibinfo{title}{Measuring complexity in an aquatic ecosystem}.
\newblock In \bibinfo{editor}{Castillo, L.~F.}, \bibinfo{editor}{Cristancho, M.}, \bibinfo{editor}{Isaza, G.}, \bibinfo{editor}{Pinz{\'o}n, A.} \& \bibinfo{editor}{Rodr{\'i}guez, J. M.~C.} (eds.) \emph{\bibinfo{booktitle}{Advances in Computational Biology}}, \bibinfo{pages}{83--89} (\bibinfo{publisher}{Springer International Publishing}, \bibinfo{address}{Cham}, \bibinfo{year}{2014}).

\bibitem{lopez2018forest}
\bibinfo{author}{L{\'o}pez-Rivera, J.~A.}, \bibinfo{author}{Rivera, A.~L.} \& \bibinfo{author}{Frank, A.}
\newblock \bibinfo{title}{Forest complexity in the green tonality of satellite images}.
\newblock In \emph{\bibinfo{booktitle}{Unifying Themes in Complex Systems IX: Proceedings of the Ninth International Conference on Complex Systems 9}}, \bibinfo{pages}{184--188} (\bibinfo{organization}{Springer}, \bibinfo{year}{2018}).

\bibitem{lopez2023temporal}
\bibinfo{author}{L{\'o}pez-D{\'\i}az, A.~J.}, \bibinfo{author}{S{\'a}nchez-Puig, F.} \& \bibinfo{author}{Gershenson, C.}
\newblock \bibinfo{journal}{\bibinfo{title}{Temporal, structural, and functional heterogeneities extend criticality and antifragility in random boolean networks}}.
\newblock {\emph{\JournalTitle{Entropy}}} \textbf{\bibinfo{volume}{25}}, \bibinfo{pages}{254} (\bibinfo{year}{2023}).

\bibitem{makridis2020formalizing}
\bibinfo{author}{Makridis, M.}, \bibinfo{author}{Leclercq, L.}, \bibinfo{author}{Ciuffo, B.}, \bibinfo{author}{Fontaras, G.} \& \bibinfo{author}{Mattas, K.}
\newblock \bibinfo{journal}{\bibinfo{title}{Formalizing the heterogeneity of the vehicle-driver system to reproduce traffic oscillations}}.
\newblock {\emph{\JournalTitle{Transportation Research Part C: Emerging Technologies}}} \textbf{\bibinfo{volume}{120}}, \bibinfo{pages}{102803} (\bibinfo{year}{2020}).

\bibitem{axenie2022antifragileoncology}
\bibinfo{author}{Axenie, C.}, \bibinfo{author}{Kurz, D.} \& \bibinfo{author}{Saveriano, M.}
\newblock \bibinfo{journal}{\bibinfo{title}{Antifragile control systems: The case of an anti-symmetric network model of the tumor-immune-drug interactions}}.
\newblock {\emph{\JournalTitle{Symmetry}}} \textbf{\bibinfo{volume}{14}}, \bibinfo{pages}{2034} (\bibinfo{year}{2022}).

\bibitem{angeli2004detection}
\bibinfo{author}{Angeli, D.}, \bibinfo{author}{Ferrell~Jr, J.~E.} \& \bibinfo{author}{Sontag, E.~D.}
\newblock \bibinfo{journal}{\bibinfo{title}{Detection of multistability, bifurcations, and hysteresis in a large class of biological positive-feedback systems}}.
\newblock {\emph{\JournalTitle{Proceedings of the National Academy of Sciences}}} \textbf{\bibinfo{volume}{101}}, \bibinfo{pages}{1822--1827} (\bibinfo{year}{2004}).

\bibitem{hizanidis2008delay}
\bibinfo{author}{Hizanidis, J.}, \bibinfo{author}{Aust, R.} \& \bibinfo{author}{Sch{\"o}ll, E.}
\newblock \bibinfo{journal}{\bibinfo{title}{Delay-induced multistability near a global bifurcation}}.
\newblock {\emph{\JournalTitle{International Journal of Bifurcation and Chaos}}} \textbf{\bibinfo{volume}{18}}, \bibinfo{pages}{1759--1765} (\bibinfo{year}{2008}).

\bibitem{lopez2019rise}
\bibinfo{author}{L{\'o}pez-Corona, O.}, \bibinfo{author}{Ramrez-Carrillo, E.} \& \bibinfo{author}{Magallanes-Guij{\'o}n, G.}
\newblock \bibinfo{journal}{\bibinfo{title}{The rise of the technobionts: toward a new ontology to understand current planetary crisis}}.
\newblock {\emph{\JournalTitle{RESEARCHERS. ONE. Available at https://www. researchers. one/article/2019-01-1}}}  (\bibinfo{year}{2019}).

\bibitem{bayer2023games}
\bibinfo{author}{Bayer, P.} \& \bibinfo{author}{West, J.}
\newblock \bibinfo{journal}{\bibinfo{title}{Games and the treatment convexity of cancer}}.
\newblock {\emph{\JournalTitle{Dynamic Games and Applications}}}  (\bibinfo{year}{2023}).

\bibitem{crosato2018critical}
\bibinfo{author}{Crosato, E.}, \bibinfo{author}{Nigmatullin, R.} \& \bibinfo{author}{Prokopenko, M.}
\newblock \bibinfo{journal}{\bibinfo{title}{On critical dynamics and thermodynamic efficiency of urban transformations}}.
\newblock {\emph{\JournalTitle{Royal Society Open Science}}} \textbf{\bibinfo{volume}{5}}, \bibinfo{pages}{180863} (\bibinfo{year}{2018}).

\bibitem{gershenson2012complexity}
\bibinfo{author}{Gershenson, C.} \& \bibinfo{author}{Fern{\'a}ndez, N.}
\newblock \bibinfo{journal}{\bibinfo{title}{Complexity and information: Measuring emergence, self-organization, and homeostasis at multiple scales}}.
\newblock {\emph{\JournalTitle{Complexity}}} \textbf{\bibinfo{volume}{18}}, \bibinfo{pages}{29--44} (\bibinfo{year}{2012}).

\bibitem{kalloniatis2018fisher}
\bibinfo{author}{Kalloniatis, A.~C.}, \bibinfo{author}{Zuparic, M.~L.} \& \bibinfo{author}{Prokopenko, M.}
\newblock \bibinfo{journal}{\bibinfo{title}{Fisher information and criticality in the kuramoto model of nonidentical oscillators}}.
\newblock {\emph{\JournalTitle{Physical Review E}}} \textbf{\bibinfo{volume}{98}}, \bibinfo{pages}{022302} (\bibinfo{year}{2018}).

\bibitem{lopez2019fisher}
\bibinfo{author}{L{\'o}pez-Corona, O.} \& \bibinfo{author}{Padilla, P.}
\newblock \bibinfo{journal}{\bibinfo{title}{Fisher information as unifying concept for criticality and antifragility, a primer hypothesis}}.
\newblock {\emph{\JournalTitle{Researchers. One}}}  (\bibinfo{year}{2019}).

\bibitem{pineda2019novel}
\bibinfo{author}{Pineda, O.~K.}, \bibinfo{author}{Kim, H.}, \bibinfo{author}{Gershenson, C.} \emph{et~al.}
\newblock \bibinfo{journal}{\bibinfo{title}{A novel antifragility measure based on satisfaction and its application to random and biological boolean networks}}.
\newblock {\emph{\JournalTitle{Complexity}}} \textbf{\bibinfo{volume}{2019}} (\bibinfo{year}{2019}).

\bibitem{cannon1929organization}
\bibinfo{author}{Cannon, W.~B.}
\newblock \bibinfo{journal}{\bibinfo{title}{Organization for physiological homeostasis}}.
\newblock {\emph{\JournalTitle{Physiological Reviews}}} \textbf{\bibinfo{volume}{9}}, \bibinfo{pages}{399--431} (\bibinfo{year}{1929}).

\bibitem{nichol2016stochasticity}
\bibinfo{author}{Nichol, D.}, \bibinfo{author}{Robertson-Tessi, M.}, \bibinfo{author}{Jeavons, P.} \& \bibinfo{author}{Anderson, A.~R.}
\newblock \bibinfo{journal}{\bibinfo{title}{Stochasticity in the genotype-phenotype map: implications for the robustness and persistence of bet-hedging}}.
\newblock {\emph{\JournalTitle{Genetics}}} \textbf{\bibinfo{volume}{204}}, \bibinfo{pages}{1523--1539} (\bibinfo{year}{2016}).

\bibitem{wang2019path}
\bibinfo{author}{Wang, H.}, \bibinfo{author}{Liu, B.}, \bibinfo{author}{Ping, X.} \& \bibinfo{author}{An, Q.}
\newblock \bibinfo{journal}{\bibinfo{title}{Path tracking control for autonomous vehicles based on an improved mpc}}.
\newblock {\emph{\JournalTitle{IEEE Access}}} \textbf{\bibinfo{volume}{7}}, \bibinfo{pages}{161064--161073} (\bibinfo{year}{2019}).

\bibitem{solea2007trajectory}
\bibinfo{author}{Solea, R.} \& \bibinfo{author}{Nunes, U.}
\newblock \bibinfo{journal}{\bibinfo{title}{Trajectory planning and sliding-mode control based trajectory-tracking for cybercars}}.
\newblock {\emph{\JournalTitle{Integrated Computer-Aided Engineering}}} \textbf{\bibinfo{volume}{14}}, \bibinfo{pages}{33--47} (\bibinfo{year}{2007}).

\bibitem{antonelli2007fuzzy}
\bibinfo{author}{Antonelli, G.}, \bibinfo{author}{Chiaverini, S.} \& \bibinfo{author}{Fusco, G.}
\newblock \bibinfo{journal}{\bibinfo{title}{A fuzzy-logic-based approach for mobile robot path tracking}}.
\newblock {\emph{\JournalTitle{IEEE transactions on fuzzy systems}}} \textbf{\bibinfo{volume}{15}}, \bibinfo{pages}{211--221} (\bibinfo{year}{2007}).

\bibitem{geroliminis2012optimal}
\bibinfo{author}{Geroliminis, N.}, \bibinfo{author}{Haddad, J.} \& \bibinfo{author}{Ramezani, M.}
\newblock \bibinfo{journal}{\bibinfo{title}{Optimal perimeter control for two urban regions with macroscopic fundamental diagrams: A model predictive approach}}.
\newblock {\emph{\JournalTitle{IEEE Transactions on Intelligent Transportation Systems}}} \textbf{\bibinfo{volume}{14}}, \bibinfo{pages}{348--359} (\bibinfo{year}{2012}).

\bibitem{zhou2023scalable}
\bibinfo{author}{Zhou, D.} \& \bibinfo{author}{Gayah, V.~V.}
\newblock \bibinfo{journal}{\bibinfo{title}{Scalable multi-region perimeter metering control for urban networks: A multi-agent deep reinforcement learning approach}}.
\newblock {\emph{\JournalTitle{Transportation Research Part C: Emerging Technologies}}} \textbf{\bibinfo{volume}{148}}, \bibinfo{pages}{104033} (\bibinfo{year}{2023}).

\bibitem{pisarchik2014control}
\bibinfo{author}{Pisarchik, A.~N.} \& \bibinfo{author}{Feudel, U.}
\newblock \bibinfo{journal}{\bibinfo{title}{Control of multistability}}.
\newblock {\emph{\JournalTitle{Physics Reports}}} \textbf{\bibinfo{volume}{540}}, \bibinfo{pages}{167--218} (\bibinfo{year}{2014}).

\bibitem{grziwotz2023anticipating}
\bibinfo{author}{Grziwotz, F.} \emph{et~al.}
\newblock \bibinfo{journal}{\bibinfo{title}{Anticipating the occurrence and type of critical transitions}}.
\newblock {\emph{\JournalTitle{Science Advances}}} \textbf{\bibinfo{volume}{9}}, \bibinfo{pages}{eabq4558} (\bibinfo{year}{2023}).

\bibitem{gatenby2009adaptive}
\bibinfo{author}{Gatenby, R.~A.}, \bibinfo{author}{Silva, A.~S.}, \bibinfo{author}{Gillies, R.~J.} \& \bibinfo{author}{Frieden, B.~R.}
\newblock \bibinfo{journal}{\bibinfo{title}{Adaptive therapy}}.
\newblock {\emph{\JournalTitle{Cancer Research}}} \textbf{\bibinfo{volume}{69}}, \bibinfo{pages}{4894--4903} (\bibinfo{year}{2009}).

\bibitem{read2011evolution}
\bibinfo{author}{Read, A.~F.}, \bibinfo{author}{Day, T.} \& \bibinfo{author}{Huijben, S.}
\newblock \bibinfo{journal}{\bibinfo{title}{The evolution of drug resistance and the curious orthodoxy of aggressive chemotherapy}}.
\newblock {\emph{\JournalTitle{Proceedings of the National Academy of Sciences}}} \textbf{\bibinfo{volume}{108}}, \bibinfo{pages}{10871--10877} (\bibinfo{year}{2011}).

\bibitem{read2014antibiotic}
\bibinfo{author}{Read, A.~F.} \& \bibinfo{author}{Woods, R.~J.}
\newblock \bibinfo{journal}{\bibinfo{title}{Antibiotic resistance management}}.
\newblock {\emph{\JournalTitle{Evolution, Medicine, and Public Health}}} \textbf{\bibinfo{volume}{2014}}, \bibinfo{pages}{147} (\bibinfo{year}{2014}).

\bibitem{whelan2020resistance}
\bibinfo{author}{Whelan, C.~J.} \& \bibinfo{author}{Cunningham, J.~J.}
\newblock \bibinfo{journal}{\bibinfo{title}{Resistance is not the end: lessons from pest management}}.
\newblock {\emph{\JournalTitle{Cancer Control}}} \textbf{\bibinfo{volume}{27}}, \bibinfo{pages}{1073274820922543} (\bibinfo{year}{2020}).

\bibitem{cunningham2019call}
\bibinfo{author}{Cunningham, J.~J.}
\newblock \bibinfo{journal}{\bibinfo{title}{A call for integrated metastatic management}}.
\newblock {\emph{\JournalTitle{Nature Ecology \& Evolution}}} \textbf{\bibinfo{volume}{3}}, \bibinfo{pages}{996--998} (\bibinfo{year}{2019}).

\bibitem{zhang2017integrating}
\bibinfo{author}{Zhang, J.}, \bibinfo{author}{Cunningham, J.~J.}, \bibinfo{author}{Brown, J.~S.} \& \bibinfo{author}{Gatenby, R.~A.}
\newblock \bibinfo{journal}{\bibinfo{title}{Integrating evolutionary dynamics into treatment of metastatic castrate-resistant prostate cancer}}.
\newblock {\emph{\JournalTitle{Nature Communications}}} \textbf{\bibinfo{volume}{8}}, \bibinfo{pages}{1816} (\bibinfo{year}{2017}).

\bibitem{west2023fundamentals}
\bibinfo{author}{West, J.}, \bibinfo{author}{Gallaher, J.}, \bibinfo{author}{Strobl, M.}, \bibinfo{author}{Robertson-Tessi, M.} \& \bibinfo{author}{Anderson, A.~R.}
\newblock \bibinfo{title}{The fundamentals of evolutionary therapy in cancer}.
\newblock In \emph{\bibinfo{booktitle}{Cancer Systems Biology and Translational Mathematical Oncology}} (\bibinfo{publisher}{Oxford University Press}, \bibinfo{year}{2023}).

\bibitem{strobl2023treatment}
\bibinfo{author}{Strobl, M.}, \bibinfo{author}{Gallaher, J.}, \bibinfo{author}{Robertson-Tessi, M.}, \bibinfo{author}{West, J.} \& \bibinfo{author}{Anderson, A.}
\newblock \bibinfo{journal}{\bibinfo{title}{Treatment of evolving cancers will require dynamic decision support}}.
\newblock {\emph{\JournalTitle{Annals of Oncology}}} \textbf{\bibinfo{volume}{34}}, \bibinfo{pages}{867--884} (\bibinfo{year}{2023}).

\bibitem{west2019multidrug}
\bibinfo{author}{West, J.~B.} \emph{et~al.}
\newblock \bibinfo{journal}{\bibinfo{title}{Multidrug cancer therapy in metastatic castrate-resistant prostate cancer: An evolution-based strategy}}.
\newblock {\emph{\JournalTitle{Clinical Cancer Research}}} \bibinfo{pages}{clincanres--0006} (\bibinfo{year}{2019}).

\bibitem{ripple2012trophic}
\bibinfo{author}{Ripple, W.~J.} \& \bibinfo{author}{Beschta, R.~L.}
\newblock \bibinfo{journal}{\bibinfo{title}{Trophic cascades in yellowstone: the first 15 years after wolf reintroduction}}.
\newblock {\emph{\JournalTitle{Biological Conservation}}} \textbf{\bibinfo{volume}{145}}, \bibinfo{pages}{205--213} (\bibinfo{year}{2012}).

\bibitem{beschta2016riparian}
\bibinfo{author}{Beschta, R.~L.} \& \bibinfo{author}{Ripple, W.~J.}
\newblock \bibinfo{journal}{\bibinfo{title}{Riparian vegetation recovery in yellowstone: the first two decades after wolf reintroduction}}.
\newblock {\emph{\JournalTitle{Biological Conservation}}} \textbf{\bibinfo{volume}{198}}, \bibinfo{pages}{93--103} (\bibinfo{year}{2016}).

\bibitem{wright2002ecosystem}
\bibinfo{author}{Wright, J.~P.}, \bibinfo{author}{Jones, C.~G.} \& \bibinfo{author}{Flecker, A.~S.}
\newblock \bibinfo{journal}{\bibinfo{title}{An ecosystem engineer, the beaver, increases species richness at the landscape scale}}.
\newblock {\emph{\JournalTitle{Oecologia}}} \textbf{\bibinfo{volume}{132}}, \bibinfo{pages}{96--101} (\bibinfo{year}{2002}).

\bibitem{ramirez2023similar}
\bibinfo{author}{Ram{\'\i}rez-Carrillo, E.} \emph{et~al.}
\newblock \bibinfo{journal}{\bibinfo{title}{Similar connectivity of gut microbiota and brain activity networks is mediated by animal protein and lipid intake in children from a mexican indigenous population}}.
\newblock {\emph{\JournalTitle{PLoS One}}} \textbf{\bibinfo{volume}{18}}, \bibinfo{pages}{e0281385} (\bibinfo{year}{2023}).

\bibitem{isaac2023potential}
\bibinfo{author}{Isaac, G.}, \bibinfo{author}{Ram{\'\i}rez-Carrillo, E.}, \bibinfo{author}{Sanchez, J.~D.}, \bibinfo{author}{L{\'o}pez-Corona, O.} \emph{et~al.}
\newblock \bibinfo{journal}{\bibinfo{title}{Potential long consequences from internal and external ecology: loss of gut microbiota antifragility in children from an industrialized population compared with an indigenous rural lifestyle}}.
\newblock {\emph{\JournalTitle{Journal of Developmental Origins of Health and Disease}}} \bibinfo{pages}{1--12} (\bibinfo{year}{2023}).

\end{thebibliography}

\end{document}